\documentstyle[sprocl,psfig]{article}

\def\Journal#1#2#3#4{{#1} {\bf #2}, #3 (#4)}

\def\PLB{{\em Phys. Lett.}  B}

\def\PRD{{\em Phys. Rev.} D}
\def\ZPC{{\em Z. Phys.} C}

\def\ra{\rightarrow}

\def\be{\begin{equation}}
\def\ee{\end{equation}}
\def\bea{\begin{eqnarray}}
\def\eea{\end{eqnarray}}

\begin{document}
\begin{flushright}
JLAB--THY--96--01

August 6, 1996
\vskip 1 cm
\end{flushright}

\title{VIRTUAL COMPTON SCATTERING ON THE PROTON\\
AT HIGH $s$ AND LOW $t$ \footnote{Invited talk at the Workshop on
Virtual Compton Scattering, Clermont--Ferrand, France, June 26--29, 1996}}

\author{Andrei Afanasev \footnote{Also at Theory Division, Kharkov
Institute of Physics and Technology, Kharkov, Ukraine}}

\address{ Jefferson Lab,
12000 Jefferson Ave, Newport News, VA 23606, USA\\ 
and\\
Department of Physics, Hampton University, Hampton, VA 23668, USA} 

\maketitle\abstracts{Virtual Compton Scattering (VCS) at low transferred 
momenta to
the proton ($t$) and sufficiently high c.m. energies ($s$) may be used to
a) study $Q^2$--dependence of leading $t$--channel exchanges and b) 
look for onset of scaling behavior with increasing $Q^2$. I discuss the 
implications for perturbative and nonperturbative QCD and suggest 
possible experiments.}

\section*{Introduction}

Analysing various kinematic domains of Virtual Compton Scattering (VCS),
one may obtain clues to various problems of strong interaction dynamics.

If $s\gg-t$, and $Q^2$ is fixed, the amplitude of VSC may be described
(Fig.1a) by the sum of $t$--channel Reggeon (R) exchanges \cite{dg},
\be {\cal M}=\sum_{R} s^{\alpha_R(t)} \beta_R(t,Q^2) e^{-{1\over 2}i 
\pi \alpha_R(t)},\ee
where the sum is taken over all possible Reggeons  with positive
charge parity, $i.e.$ Pomeron ($\alpha(0)\approx 1$), $f-a_2$
($\alpha(0)\approx 1/2$) and pion ($\alpha(0)\approx 0$) trajectories.
It implies that at asymptotically high energies ($s\to\infty$), the energy
behavior of Compton amplitude is governed by the Pomeron exchange. 

However, as I demonstrate below, if $s$ and $Q^2$ are in the range of a 
few (GeV/c)$^2$, and $t\approx-m_\pi^2$, where $m_\pi$ is a pion mass,
a relative contribution from the $\pi^0$--exchange becomes
large, exceeding even the diffractive (Pomeron) contribution at low $t$.
Taking advantage of different quantum numbers and phases of $t$--channel
exchanges, it is possible to separate these contributions \cite{afanas94}
and measure the form factor of $\gamma^*\pi^0\to\gamma$ transition 
($F_{\gamma^*\gamma\pi^0}(Q^2)$) as a 
function of $Q^2$.  Theoretically, the 
leading--twist QCD contribution to the 
$F_{\gamma^*\gamma\pi^0}$ transition form factor is given by the quark
triangle diagram (Fig.1b) with no hard gluon exchange making this process 
0th--order in QCD running coupling constant
$\alpha_s$ 
with non--perturbative dynamics being contained in the pion 
distribution amplitude
$\varphi_\pi(x)$. \cite{bl} The function  $\varphi_\pi(x)$
cannot be predicted by perturbative QCD (except for it asymptotic behavior)
and it is crucial for 
understanding whether or not one may apply a perturbative QCD description
to exclusive processes at given energies. The data on  
$F_{\gamma^*\gamma\pi^0}(Q^2)$ obtained at $e^+e^-$ colliders 
\cite{cello,cleo} currently available for $Q^2$ up to 8 (GeV/c)$^2$ 
indicate that 
$\varphi_\pi(x)$ is close to its asymptotic form, and therefore 
`soft', nonperturbative mechanisms are dominant in this energy range,
in agreement with theoretical predictions based on QCD Sum Rules.\cite{rr}

This conclusion is so important that it is desirable to have an
independent measurement of $F_{\gamma^*\gamma\pi^0}(Q^2)$,
and such a measurement via VCS was proposed earlier at 
Jefferson Lab,\cite{loi} for transferred momenta 
$Q^2=1.0\div 3.5$ (GeV/c)$^2$.
Higher values of $Q^2$ are also possible for 
the Jefferson Lab energy upgrade.

As $Q^2$ increases, and $s$ stays large, in the low--$t$ limit  one may 
observe scaling behavior of the VCS amplitude predicted recently by 
X.~Ji \cite{ji} and A.~Radyushkin. \cite{deeply} 
 
\begin{figure}
\hskip 1. in\psfig{figure=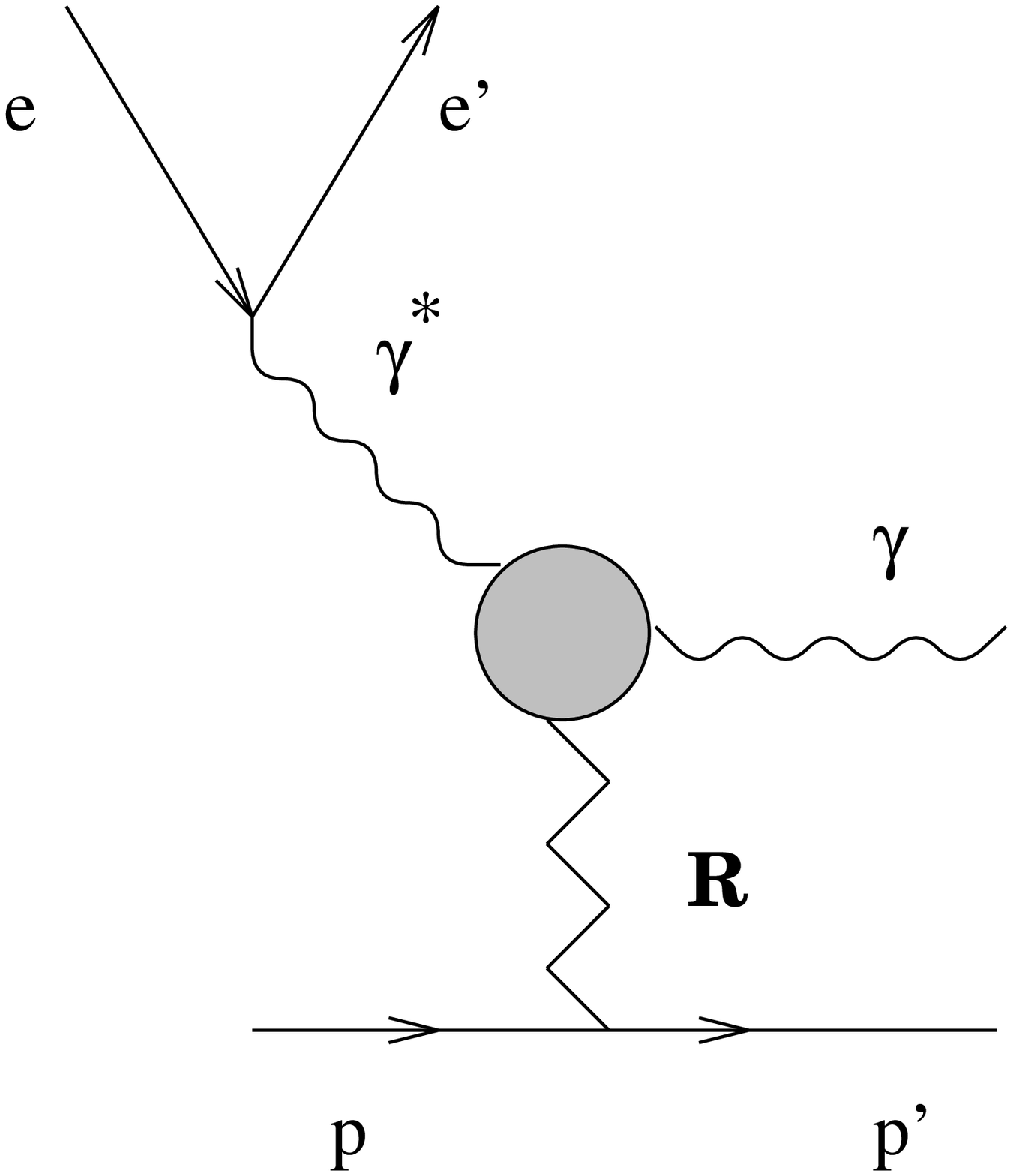,height=1.5 in}
\vskip -1.9 in
\hskip 3. in\psfig{figure=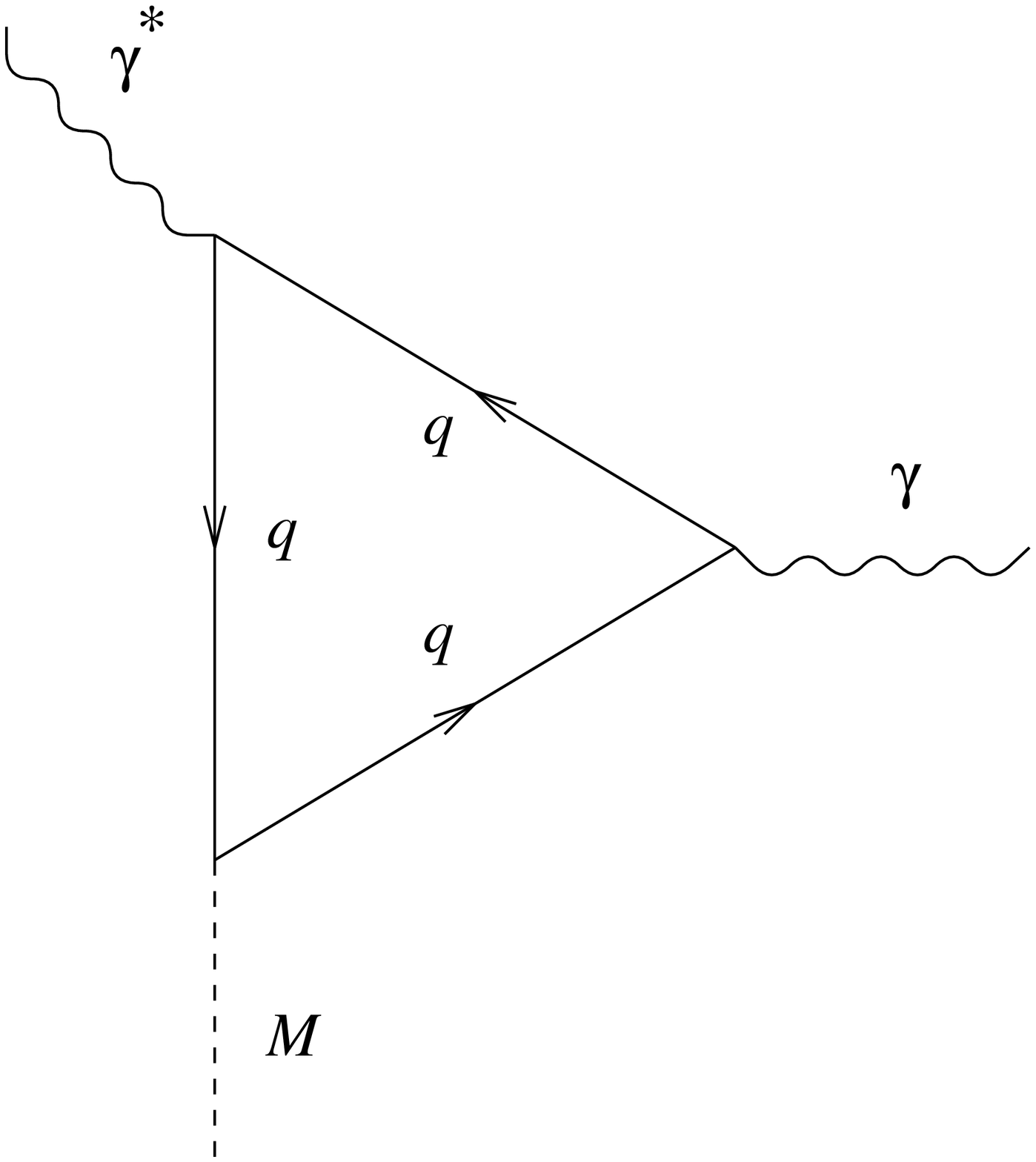,height=1.5 in}

\hskip 1.3 in a)
\hskip 1.5 in b)
\caption{a) VSC amplitude in terms of $t$--channel Reggeon exchanges, where
$R$ stands for the Pomeron, $f-a_2$, and pion trajectories, and the dashed 
blob denotes the two--photon transition form factor; 
b) Quark triangle diagram for $\gamma^*M\to\gamma$ transition, where
$M$ is a pseudoscalar meson.}
\label{fig:regge}
\end{figure}

\section*{Transition Form Factors of Leading $t$--channel Exchanges}

Regge phenomenology proved very successful at describing energy dependence
of hadronic total cross sections and differential
cross--sections at low $t$ and high $s$.  For VCS, it predicts the slope
of the Regge trajectory $\alpha_R(t)$ to be independent of the photon
virtuality $Q^2$. This is an important result of Regge theory which needs
to be tested experimentally, as was indicated earlier (see, $e.g.$, 
Ref. \cite{brodsky}). 

As far as form factors of $\gamma^* Reggeon\ra\gamma$ transitions are 
concerned, one needs a microscopic theory of Reggeon exchanges to be able
to predict their $Q^2$--dependence. This task is challenging
for the `soft' Pomeron exchange limited to a few GeV energy scale, and
this problem is still far from being solved in QCD. 

However, QCD predictions for the two--photon transition 
form factors
related to the $t$--channel of VCS are available for the case of
$\pi^0$ exchange. For the review of theoretical approaches,  
see Ref.\cite{rr} I would also like to mention here effective quark models for 
$F_{\gamma^*\gamma\pi^0}$ based on the extended NJL--model \cite{ibg} and
the model of dynamical dressing of propagators and vertices,\cite{offshell} 
the latter also addressed
the off-shell behavior of $F_{\gamma^*\gamma\pi^0}$ for the proposed Jefferson
Lab experiment. \cite{loi}

The lightest meson which can
be exchanged in the $t$--channel of VCS is $\pi^0$. It may be possible
to extract
the corresponding form factor doing a Chew--Low extrapolation like
for the charged pion form factor 
measurements \cite{mack} scheduled at Jefferson Lab. 

\section*{Exchange of $\pi^0$}

In the real Compton scattering on the proton at high $s$ and low $t$, the
main contribution to the cross section is known to be diffractive, due to
the Pomeron exchange.
I will demonstrate here that the situation is different for the
case of virtual photons, because of strong enhancement of 
the $\pi^0$--exchange term.

At small
$t$, the  $\pi$--trajectory is determined by the pion Born term.
The matrix element of the corresponding transition is
\bea
{\cal M}_{\gamma^*p\to\gamma p}^{(\pi^0)} = e^2 F_{\gamma^*\gamma\pi^0}(Q^2)
 g_{\pi NN} F_{\pi NN}(t) 
D_\pi(t)\epsilon_{\mu\nu\alpha\beta}\varepsilon_\mu\varepsilon_\nu'^{*}
q_\alpha q_\beta ' {\bar u'}\gamma_5 u,
\eea
where $\varepsilon (\varepsilon')$ is the polarization 4--vector of initial
(final) photon, and $q (q')$ is its momentum ($Q^2\equiv-q^2$), and
$u(u')$ is a bispinor of the initial (final) proton. The pion propagator 
has the form $i D_\pi(t)=(t-m_\pi^2)^{-1}$, and I also assumed a conventional
monopole form for the cut--off form factor 
$F_{\pi NN}(t)=\Lambda^2/(\Lambda^2-t)$.

Define four coincidence structure functions (SF)  for the (unpolarized)
$p(e,e'\gamma)p$ cross section as
\bea
{d^5\sigma\over d E' d\Omega_e d \Omega_p}&=&{\alpha^3\over 16\pi^3}{E'\over E}
{|{\bf p}|_{c.m.}\over m W}{1\over Q^2} {1\over 1-\epsilon} 
[\sigma_T+\epsilon \sigma_L+ \epsilon \cos(2\varphi) \sigma_{TT}\nonumber \\
& &+\sqrt{2\epsilon (1+\epsilon)} \cos(\varphi) \sigma_{LT}],\\
\epsilon^{-1}&\equiv &1-2{{\bf q}_{lab}^2\over q^2} \tan^2{\theta_e\over 2},
\eea
where $E(E')$ is the initial (final) electron energy, $\theta_e$ is the
electron scattering angle, $\varphi$ is the 
azimuthal angle, and $m$ is the proton mass. The matrix element given by eq.(2)
yields the following contributions to SF:
\bea
\sigma_T^{(\pi^0)}&=&[(|{\bf q}|-q_0\cos\theta)^2+
(|{\bf q}|\cos\theta-q_0)^2] X, \\
\sigma_L^{(\pi^0)}&=& 2 Q^2 \sin^2(\theta) X, \\
\sigma_{LT}^{(\pi^0)}&=&2 \sqrt{Q^2} (|{\bf q}|-q_0\cos\theta)\sin(\theta) X,\\
\sigma_{TT}^{(\pi^0)}&=& Q^2 \sin^2(\theta) X, \\
X&=&{-t\over (t-m_\pi^2)^2} [q_0' F_{\pi NN}(t) g_{\pi NN} 
F_{\gamma^*\gamma\pi^0}(Q^2)]^2,
\eea
where the c.m. energy of the (initial) final photon is given 
by $q_0={q^2+W^2-m^2\over 2 W}$ ($q'_0={W^2-m^2\over 2 W}$),
and $\theta$ is the c.m. angle of outgoing photon. 

\begin{figure}[t]
\vskip -.7 in
\hskip 1.5 in \psfig{figure=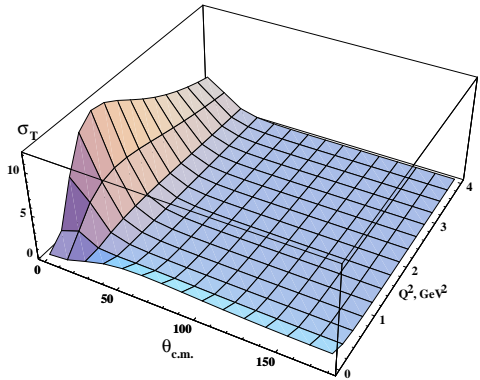,height=3.5 in}
\vskip -.7 in

\caption{Contribution of the $\pi^0$--exchange Born diagram to the 
(dimensionless) structure function $\sigma_T$ of transverse virtual 
photoabsorption; the invariant mass of final $\gamma p$ state is
taken $W$= 2.5 GeV.
\label{fig:sigmat}}
\vskip -.2 in
\end{figure}

The $\pi^0$ pole contributes the most to
transverse photoabsorption in VCS; the corresponding SF
is shown in Fig.~\ref{fig:sigmat}. Contributions to the other structure 
functions are suppressed
at small $t$ by the factor of $\theta$ for $\sigma_{LT}$ and $\theta^2$ for
$\sigma_{L}$ and $\sigma_{TT}$.
The  overall factor $-t/(t-m_\pi^2)^2$ from Eq.(9) has a pole 
in the unphysical
region, $t=m_\pi^2$, turns to zero at $t=0$, and has a maximum in the
physical region at $t=-m_\pi^2$. As can be seen from Fig.~\ref{fig:sigmat},
dependence of the $\pi^0$ contribution on $\theta$
changes dramatically  when
going from real to virtual photons. When $Q^2=0$, it is $suppressed$ at
forward angles ($i.e.$, low $t$); in contrast, for $Q^2\neq 0$, it is 
$peaked$ at forward angles. This result is due to the Lorentz structure of 
the $\gamma^*\gamma\pi^0$--vertex defined by Eq.(2): 
 $\epsilon_{\mu\nu\alpha\beta}\varepsilon_\mu\varepsilon_\nu'^{*} 
q_\alpha q_\beta '$.  

Enhancement of $\pi^0$ exchange makes it large enough to reach,
and even exceed, the magnitude of diffractive term. For instance,
assuming the VMD--model for the $\gamma^*Pomeron\ra\gamma$ transition form
factor, $\pi^0$ contribution to $\sigma_T$ is evaluated to be 
25\% higher than from the
Pomeron at $Q^2=$ 2 (GeV/c)$^2$ and, respectively, three times higher at
$Q^2=$ 3 (GeV/c)$^2$. (This result is obtained at $W=$ 2.5 GeV, $t=-m_\pi^2$).

It creates favorable conditions for extracting the form factor 
$F_{\gamma^*\gamma\pi^0}$ from VCS experiments. 

\section*{Separation of the $t$--channel exchanges}

One can attempt to disentangle various $t$--channel exchanges in VCS.

Indeed, the Pomeron a) has quantum numbers of vacuum (except for its spin),
b) contributes almost purely to the imaginary part of VCS amplitude at low
$t$, and c) does not flip the nucleon spin. On the other hand, the pion
a) is isovector and pseudoscalar b) contributes to the real part of
VCS amplitude at low $t$, and c) flips the nucleon spin. 

These circumstances may be used to design the experiments in 
order to separate these mechanisms, especially if 
nuclear targets are used. For instance,
coherent VCS on $^4$He  excludes pseudoscalar $t$--channel exchanges 
($\pi^0,\eta,\eta'$), thus providing useful information on $Q^2$ 
evolution of the diffractive (Pomeron) term. Coherent VCS on deuterium target 
would rule out the $\pi^0$--exchange, but keep exchange of other pseudoscalars
with zero isospin. On the other hand, if the hadronic target undergoes an 
isovector transition of any kind ($e.g.$ threshold deuteron dissociation),
it would be completely due to the $\pi^0$ exchange.

In addition,
both diffractive and pseudoscalar $t$--channel exchanges are strongly 
suppressed in interference between VCS and the Bethe--Heitler amplitudes 
for the case of unpolarized particles. If, however, spin effects are
included, the asymmetry and/or recoil polarization due to the
proton polarized normal to the reaction plane would be caused by interference
between diffractive and $\pi^0$ terms (in which case the electron beam
polarization is not required), while the sideways (in--plane) 
asymmetry/polarization with longitudinally polarized electrons would be 
mainly caused by interference between
$\pi^0$--exchange and the Bethe--Heitler amplitudes. 

\section*{Summary}

$\bullet$ Exchanges of $\pi^0$ and the Pomeron in $t$--channel provide the
largest contributions to the amplitude of VCS on the proton at small $t$ and
$s$ in the region of a few GeV$^2$.

$\bullet$ It is possible to separate these contributions and study 
$Q^2$--dependence of the corresponding form factors.

$\bullet$ For the form factor $F_{\gamma^*\gamma\pi^0}$, it gives information
about the pion distribution amplitude and QCD corrections. It also
discriminates between predictions of effective quark models. 

$\bullet$ Further increasing energies and momentum transfers, and keeping 
$t$ small, one may observe a transition to the scaling limit of VCS predicted
and studied theoretically by X. Ji \cite{ji} and A. Radyushkin.\cite{deeply}

\section*{Acknowledgments}
I would like to thank Vincent Breton and the other members of the organizing
committee for the perfectly organized workshop at Clermont--Ferrand. 
Discussions with A. Radyushkin, N. Isgur, S. Brodsky, and P. Bertin are 
gratefully acknowledged. This work was supported by the US Department of 
Energy under contract DE--AC05--84ER40150.

\end{document}